\documentclass[12pt]{article}

\usepackage{latexsym}
\usepackage[auth-sc,affil-sl]{authblk}
\usepackage{graphicx}
\usepackage{color}
\usepackage{amsmath}
\usepackage{dsfont}
\usepackage[round]{natbib}
\usepackage{amssymb}
\usepackage{abstract}
\usepackage{hyperref}
\usepackage[margin=1in]{geometry}
\usepackage{enumerate}
\usepackage{accents}
\usepackage{subcaption}
\usepackage{fancyvrb}

\newcommand{\qu}[1]{``#1''}

\newcounter{probnum}
\allowdisplaybreaks

\definecolor{gray}{rgb}{0.7,0.7,0.7}

\definecolor{black}{rgb}{0,0,0}
\definecolor{white}{rgb}{1,1,1}
\definecolor{blue}{rgb}{0,0,0.7}

\definecolor{green}{rgb}{0.133,0.545,0.133}

\definecolor{yellow}{rgb}{1,0.549,0}

\definecolor{red}{rgb}{1,0.133,0.133}

\definecolor{purple}{rgb}{0.58,0,0.827}

\definecolor{brown}{rgb}{0.55,0.27,0.07}

\definecolor{backgcode}{rgb}{0.97,0.97,0.8}
\definecolor{Brown}{cmyk}{0,0.81,1,0.60}
\definecolor{OliveGreen}{cmyk}{0.64,0,0.95,0.40}
\definecolor{CadetBlue}{cmyk}{0.62,0.57,0.23,0}

\newcommand{\bv}[1]{\boldsymbol{#1}}

\newcommand{\sigsq}{\sigma^2}

\newcommand{\thetahat}{\hat{\theta}}
\newlength{\dhatheight}
\newcommand{\doublehat}[1]{%
    \settoheight{\dhatheight}{\ensuremath{\hat{#1}}}%
    \addtolength{\dhatheight}{-0.35ex}%
    \hat{\vphantom{\rule{1pt}{\dhatheight}}%
    \smash{\hat{#1}}}}
\newcommand{\thetahathat}{\doublehat{\theta}}
\newcommand{\varthetahat}{\mathbb{V}\text{ar}[\thetahat]}
\newcommand{\varthetahathat}{\mathbb{V}\text{ar}[\thetahathat]}

\newcommand{\xbar}{\bar{x}}

\newcommand{\xmin}{x_{\text{min}}}
\newcommand{\xmax}{x_{\text{max}}}

\newcommand{\iid}{~{\buildrel iid \over \sim}~}
\newcommand{\inddist}{~{\buildrel ind \over \sim}~}

\newcommand{\B}{\bv{B}}

\newcommand{\X}{\bv{X}}

\newcommand{\Y}{\bv{Y}}

\newcommand{\Xt}{\bv{X}^\top}
\newcommand{\XtX}{\Xt\X}
\newcommand{\XtXinv}{\parens{\Xt\X}^{-1}}

\newcommand{\bbeta}{\bv{\beta}}
\newcommand{\bOmega}{\bv{\Omega}}

\newcommand{\twovec}[2]{\bracks{\begin{array}{c} #1 \\ #2 \end{array}}}

\newcommand{\beqn}{\vspace{-0.25cm}\begin{eqnarray*}}
\newcommand{\eeqn}{\end{eqnarray*}}
\newcommand{\bneqn}{\vspace{-0.25cm}\begin{eqnarray}}
\newcommand{\eneqn}{\end{eqnarray}}

\newcommand{\parens}[1]{\left(#1\right)}
\newcommand{\squared}[1]{\parens{#1}^2}

\newcommand{\prob}[1]{\mathbb{P}\parens{#1}}
\newcommand{\cprob}[2]{\prob{#1~|~#2}}

\newcommand{\bracks}[1]{\left[#1\right]}
\newcommand{\braces}[1]{\left\{#1\right\}}

\newcommand{\inverse}[1]{\parens{#1}^{-1}}

\newcommand{\var}[1]{\mathbb{V}\text{ar}\bracks{#1}}

\newcommand{\overtwo}[1]{\frac{#1}{2}}
\newcommand{\overn}[1]{\frac{#1}{n}}

\newcommand{\multnormnot}[3]{\mathcal{N}_{#1}\parens{#2,\,#3}}
\newcommand{\normnot}[2]{\mathcal{N}\parens{#1,\,#2}}

\newcommand{\zeroonecl}{\bracks{0,1}}

\newcommand{\convp}{~{\buildrel p \over \rightarrow}~}

\newcommand{\convd}{~{\buildrel \mathcal{D} \over \rightarrow}~}

\renewcommand{\min}[1]{\text{min}\braces{#1}}

\newcommand{\errorrv}{\mathcal{E}}
\newcommand{\berrorrv}{\bv{\errorrv}}

\title{Optimal Experimental Designs for Estimating Henry's Law Constants via the Method of Phase Ratio Variation}

\author[1]{Adam Kapelner\thanks{Electronic address: \texttt{kapelner@qc.cuny.edu}; Prinicipal Corresponding author}}
\author[2]{Abba Krieger\thanks{Electronic address: \texttt{krieger@wharton.upenn.edu}; Corresponding author}}
\author[3]{William J. Blanford\thanks{Electronic address: \texttt{william.blanford@qc.cuny.edu}; Corresponding author}}

\affil[1]{Department of Mathematics,  Queens College,  The City  \newline University of New York}
\affil[2]{Department of Statistics, The Wharton School of the \newline University of Pennsylvania}
\affil[3]{School of Earth and Environmental Sciences, Queens College,  The City  \newline University of New York}

\begin{document}
\maketitle

\begin{abstract}
When measuring Henry's Law constants ($k_H$) using the phase ratio method via headspace gas chromatography (GC), the value of $k_H$ of the compound under investigation is calculated from the ratio of the slope to the intercept of a linear regression of the the inverse GC response versus the ratio of gas to liquid volumes of a series of vials drawn from the same parent solution. Thus, an experimenter will collect measurements consisting of the independent variable (the gas/liquid volume ratio) and dependent variable (the inverse GC peak area). There is a choice of values of the independent variable during measurement. A review of the literature found that the common approach is a simple uniformly spacing of the liquid volumes. We present an optimal experimental design which estimates $k_H$ with minimum error and provides multiple means for building confidence intervals for such estimates. We illustrate efficiency improvements of our new design with an example measuring the $k_H$ for napthalene in aqueous solution as well as simulations on previous studies. The designs can be easily computed using our open source software \texttt{optDesignSlopeInt}, an \texttt{R} package on \texttt{CRAN}. We also discuss applicability of this method to other fields.
\end{abstract}
\vspace{3cm}
\pagebreak

\section{Introduction}\label{sec:intro}

\citet{Henry1803} observed that the amount of gases such as carbon dioxide, hydrogen sulfide, and others were taken up by water at a particular temperature were proportional to their partial pressures. Subsequently referred to as Henry’s Law or Henry's Constant, it is a ubiquitously used metric especially critical in chemical processing and environmental sciences. \citet{Mackay2006} among others has compiled measurements of Henry's Constant for over a thousand compounds for pure water under standard temperature and pressure as well as additional salinities and temperatures using a variety of methods.  One of those methods is called the \qu{Phase Ratio Variation Method} (PRV), and was developed by \citet{Ettre1993}. Its development marked a substantial improvement over previous methods which required external standards of known concentration. 

In brief, the PRV involves first preparing a series of subsamples from the same stock containing a dilute concentration of a volatile organic compound (VOC) solution with different ratios of gas to liquid volumes. After equilibration under the same temperature and pressure conditions, the headspace of the vials is sampled and analyzed by gas chromatography (GC). The dimensionless form of the Henry’s Constant can then be determined from the slope divided by the intercept of the inverse of the GC peak area ($G_C^{-1}$) versus the gas to liquid volume ratio (for a derivation of this relationship, see Equation 8 in \citealp{Ramachandran1996}). 

It should be noted that the PRV as developed by \citet{Ettre1993} assumes a linear response for the GC detector. This is largely true with most detectors having a linear dynamic range of three to seven orders of magnitude in concentration. At the upper and lower ends of the detector detectable range, non-linear behavior is commonly observed. \citet{Atlan2006} developed a modified version of the PRV for use in this non-linear range of GC detectors. They found the best results featured repeated analysis of an analyte over a range of concentration and for several different detectors with different relative response functions. Note that \citet{Ettre1993} also observed non-linearity in extremely volatile solutes (e.g. $k_H > 144$) and these are not explored herein.

As with with most measuring devices in this field, it is commonly necessary to analyze samples of known composition to determine standard parameters (such as solubility or vapor pressure). However, this calibration is minimal with PRV (see \citealp{Chaintreau1995} for more discussion) and mostly done to ensure proper operation of the equipment.

Herein we focus here on the method most commonly in use, linear regression. As an example of the widespread use of this method, Table~\ref{tab:previous_studies} shows example measurements by previous studies, examples which we will return to when we discuss our method in Section~\ref{sec:new_design}.

\begin{table}[htp]
\centering
\begin{tabular}{l|l|cccc}
\#  & Authors & Solutes       & Solvent  \\\hline
1& \citet{VanDurme2015}  & fifty indoor VOCs & Nalophan   \\
2& \citet{Benjamin2011}   & five VOCs & Oil-in-water emulsions   \\
3& \citet{Gao2009}          & BTEX and chlorin- & Cyclodextrin aqueous   \\
         &  &  ated solvents & solutions \\
4& \citet{Kechagia2008}   & two monomers  & Water  \\
5& \citet{Lei2007}           & alkanols & Water   \\
6& \citet{Atlan2006}        & aroma compound & Propylene glycol   \\
7& \citet{Jouquand2004}  & aroma compound & Cyclodextrin aqueous   \\
&&& solutions \\
8& \citet{Chai1998}         & methanol & Water  \\
9& \citet{Peng1998}  and         & BTEX and chlorin- & Water and saline   \\
  & \citet{Peng1997}  &  ated solvents & waters \\
10& \citet{Ettre1993}        & four VOCs & Water  \\
\hline
\end{tabular} 
\caption{Information about ten previous studies who use the PRV to measure $k_H$ for a variety of VOC's in reverse chronological order.}
\label{tab:previous_studies}
\end{table}

In essence, the volume ratio is the independent variable in the PRV method whose values are free to be chosen by the experimenter within an operating range dictated by the liquid volume (or mass) preparation and measurement devices. Thus, our problem is one of \qu{optimal design:} which volume ratios should be chosen for measuring the inverse of the GC response peak area, $G_C^{-1}$? 

Optimal design is of enormous importance and has been studied for nearly 100 years since \citet{Smith1918}'s paper on designs for polynomial regression and  \citet{Fisher1926}'s advice for treatment comparison experimentation. The field was formalized beginning with \citet{Kiefer1959}'s paper which laid the foundations for further work. Equivalence theorems for optimality criteria were to follow. Modern goals include variable screening, response surface exploration, system optimization and optimal inference \citep{Hu2014} which is our focus here. There are many good textbooks written on the subject for the interested reader e.g. \citet{Pukelsheim1993}. 

Given the heterogeneity in experimental conditions, covariate domains, parameters of interest and error structures, it is difficult to provide universal optimal designs. Thus, some work in this field focuses on tailoring designs to specific applications. An optimal design for the specific application of estimating the slope-to-intercept ratio and inference for such estimates to our knowledge has not been studied in detail (especially in the application settings of PRV and similar scientific settings) nor has specialized software been developed for this application. This is the modest goal herein.

Optimization specific to PRV is of great interest \citep{Hinshaw2006}. We would like to stress that naively estimating the slope-to-intercept ratio is dangerous: the sample slope divided by the sample intercept estimator has infinite moments and therefore can vary wildly; estimates far away from the true value are all too common in the low sample laboratory setting under modest measurement error. Simulations demonstrate our design can achieve gains one to two orders of magnitude smaller in standard error of the estimator. Since $k_H$ values are widely used in vital calculations of the phase distribution and total levels of volatile solutes at hazardous wastes and for projection of the performance of air strippers, improvements in the methods for their determination offers substantial societal benefits. 

The paper's outline is as follows. We describe our improved design for homoskedastic and heteroskedastic data in Section~\ref{sec:new_design}. We discuss many strategies for inference at the end of this section. We then illustrate an application of the optimal design by employing it to estimate the $k_H$ of napthalene in an aqueous solution in Section~\ref{sec:example} along with simulations of interval performance and robustness to a priori parameter decisions. This section also demonstrates how the software, written in \texttt{R} \citep{Rlang}, is used by an experimenter in this laboratory setting. In Section~\ref{sec:past_examples}, we estimate efficiency gains via simulation of our improved design would offer on the $k_H$ measurements in the previous studies listed in Table~\ref{tab:previous_studies}. We conclude, offer future directions as well as discuss the wider applicability of the design in Section~\ref{sec:discussion}. 

\section{An Improved Design For Measuring the Slope-to-Intercept Ratio}\label{sec:new_design}

We formalize the setup as follows. The volume ratios are the independent variable which we denote by $x$. In an experimental design, $x$ is fixed and thus is denoted with a lower case letter. Note that all $x > 0$ since volume (and therefore, volume ratios) are positive metrics. The headspace gas chromatograph (HSGC) analytical method and the test conditions (e.g. analyte solution composition and incubation temperature / pressure) in question have a range of volume ratios for which relatively high quality linear data can be assured. We denote this range $\bracks{\xmin, ~\xmax}$. Parenthetically, this range is known as the \qu{experimental domain} in the literature \citep{Pukelsheim1993}. The $G_C^{-1}$ is the dependent variable which, when measured, we denote by $y$. 

The linearity assumption from \citet{Ettre1993} helps to isolate our design problem and obviates the usual necessity to check model assumptions or employ transformations which are known to bias parameter estimates \citep{Tellinghuisen2000}. We further do not need non-linear fits which feature a host of other complications \citep{Chai1998, Tellinghuisen2000a, Atlan2006}. 

Consider $n$ runs in the experiment where $i$ indexes the run by $1, \ldots, n$. Thus, the scalar and vector relationship between $x$ and $y$ is

\bneqn
Y_i &=& \beta_0 + \beta_1 x_i + \errorrv_i \quad \text{and} \label{eq:linear_model}\\
\Y &=& \X\bbeta + \berrorrv \quad \text{respectively}\label{eq:linear_model_vec}
\eneqn

\noindent where $\X$ is the $n \times 2$ matrix with ones in the first column and the $x_i$'s in the second column, $\beta_0$ is the true $y$-intercept, $\beta_1$ is the true slope ($\bbeta$ is the vector composed of the intercept and slope), $Y$ is the random variable representation of the response ($\Y$ is its $n$-vector) and $\errorrv$ denotes the random variable representation of measurement error ($\berrorrv$ is the $n$-vector of all measurement errors) which are independent across samples and mean centered with finite variance. We now formulate the optimal design for the homoskedastic setup.

\subsection{Optimal Homoskedastic Design}\label{subsec:homo_design}

Consider the $\errorrv_i$'s (from Equation~\ref{eq:linear_model}) to be independent and identically distributed with mean zero, variance $\sigsq$ but have unknown distribution denoted by $e$:

\bneqn\label{eq:homo_errors}
\errorrv_1, \ldots, \errorrv_n \iid e(0, \sigsq).
\eneqn

\noindent Equation~\ref{eq:homo_errors} signifies we are assuming \qu{homoskedasticity} in the measurement error structure which means in the case of HSGC that no matter which volume ratio are chosen within $\bracks{\xmin,~ \xmax}$, measurements of $G_C^{-1}$ will have equal scatter from their mean value on the $k_H$ line, a restriction we relax in the next section.

The true value of $k_H$ is the statistical parameter which we denote by $\theta$ (the canonical statistical notation for the parameter of interest) and it is formed by the slope divided by the intercept, $\theta := g(\bbeta) := \beta_1 / \beta_0$. We denote $B_1$ and $B_0$ for the estimators for the slope and intercept respectively and we denote the vector $\B := \bracks{B_0~ B_1}^\top$. We then employ minimization of least squares estimation whose solution is given by

\bneqn\label{eq:ls}
\B := \inverse{\XtX} \Xt \Y 
\eneqn

\noindent and whose variance-covariance matrix is computed via

\bneqn\label{eq:ls_homo_var}
\var{\B} = \sigsq \inverse{\XtX}.
\eneqn

We consider here the natural estimator for $\theta$ which we denote $\thetahat := g(\B) = B_1 / B_0$, an estimator which converges in probability to $\theta$ (via Slutsky's Theorem) and is thereby consistent. 

We must now choose an \qu{optimality criterion} which is an approach to minimizing $\var{\B}$ of which there are many \citep[see][Chapter 9]{Pukelsheim1993}. Commonly, the literature focuses on a \qu{general} design consideration i.e. over all parameters in the model. The case we focus on is special; we only consider a univariate function of the possible parameters $g(\B) = B_1 / B_0$. 

Rather than estimating $\beta_1 / \beta_0$ by $\thetahat = B_1 / B_0$, we use a first-order Taylor approximation to produce an estimator, $ \thetahathat := g(\bbeta) + \nabla g(\bbeta)^\top \cdot (\B - \bbeta)$. It is the variance of this estimator that we choose to minimize. Note that $\bbeta$ is unknown and we estimate it by $\B$ for the purposes of implementation. Although $\varthetahat$ and $\varthetahathat$ could be very different, it follows that $\thetahathat / \thetahat \convp 1$. Hence the variance we wish to minimize is

\bneqn\label{eq:thetahat_prop_variance}
\varthetahat \approx \varthetahathat &=& \nabla g(\bbeta)^\top ~\var{\B}~ \nabla g(\bbeta) \nonumber \\
&=& \frac{\sigsq}{\beta_0^2} \bracks{-\theta ~~ 1} \XtXinv \twovec{-\theta}{1} \nonumber \\
&\propto&  \frac{\squared{\theta \xbar + 1}}{s_x^2}
\eneqn

\noindent where $\xbar$ and $s^2_x$ are the sample average and sample variance of the $x_i$'s respectively. The proportionality in the last line follows from the fact that we control only the $x_i$'s and not constants such as $n$, $\sigsq$, etc. Note that $\varthetahathat$ is a function of $\theta$, the parameter we are attempting to estimate in the first place! Thus, the experimenter must provide a rough estimate before the experiment begins. Given that numerous $k_H$ measurements and related values (e.g. vapor pressure and solubility) have been published, it is common practice to project values for unique conditions using common chemistry principles (e.g. ideal gas law, Van't Hoff equation, Raoult's law, salting out equation, etc). We discuss how robust our method is to misspecification of this estimate in Section~\ref{sec:example}. To make this distinction clear, we now notate this initial estimate as \qu{$\theta_0$} to distinguish it from the true parameter $\theta$.

Our goal is to minimize $\varthetahathat$. Recall that the experimenter has control over how the $x_i$'s are allocated within $\bracks{\xmin, \xmax}$. We refer to the allocation which minimizes this variance as the \qu{optimal design} without pretension that it is the globally optimal design. 

We argue that all of the $x_i$'s have to be a boundary point (i.e. either $\xmin$ or $\xmax$). Imagine if there were multiple points in the interior and we select two, $a$ and $b$ where $a \leq b$ where $a$ is closer to $\xmin$ than $b$ is to $\xmax$. We can move $a$ to $\xmin$ and $b$ the corresponding distance towards $\xmax$. This widening would serve to keep $\xbar$ the same and increase $s^2_x$, thereby shrinking the proportional expression for $\varthetahathat$ (Equation~\ref{eq:thetahat_prop_variance}, line 3). Repeat this procedure until there is only one possible interior point. We then prove in Appendix~\ref{app:proofs} that it is optimal for this one possible interior point to be allocated to an endpoint as well. 

We now solve for the proportion of points assigned to each endpoint in the optimal design. Assume all $x_i \in \braces{\xmin, \xmax}$ and denote the proportion of $x_i$'s equal to $\xmin$ to be $\rho$ (and thus the proportion of $x_i$'s equal to $\xmax$ will be $1 - \rho$). It is a straightforward but tedious exercise to show that the optimal proportion denoted $\rho^\star$ is given by

\bneqn\label{eq:homo_optimal_design}
\rho^\star = \frac{1 + \theta_0 \xmax}{2 + \theta_0 (\xmin + \xmax)}.
\eneqn

\noindent In the literature \citep{Kiefer1959}, $\rho^\star$ is termed the \qu{probability measure in the optimal design} and $\rho^\star n$ is termed the \qu{optimal weight} for $\xmin$. However, in any exact design, the weights must be discrete. Thus, we assign $\text{round}(\rho^\star n)$ to $\xmin$ and the remaining to $\xmax$ with the restriction that there must be at least one $x_i = \xmin$ and at least one $x_i = \xmax$. Thus, our optimal design is given via:

\bneqn\label{eq:homo_design_alg}
\# \xmin\text{'s} = \min{\max\braces{\text{round}(\rho^\star n), ~1}, n-1} ~ \text{and} ~ \# \xmax\text{'s} = n - \# \xmin\text{'s}.
\eneqn
 
\noindent We illustrate the efficiency gains of this allocation method as well as other important considerations in Section~\ref{sec:example}. 

Special designs such as the one described here must usually be counterbalanced by checking the linear model assumptions (as considered for instance by \citealp{Stigler1971}). Note that in our setup, linearity of $G_C^{-1}$ versus volume ratio need not be checked because it is a built-in assumption as long as the HSGC instrument is functioning properly. We also do not assume normality of the errors. Heteroskedasticity is addressed in the upcoming section. Thus, the standard practice of checking the linear model assumptions is not necessary when implementing our design.

In practice, there are practical limitations when using PRV based approaches to measuring $k_H$ (e.g. decreased precision and sensitivity for low volatility analytes, limited confidence in vial volume which makes volume ratios for low liquid volumes more uncertain, and extraction of significant portions of the gas volume during sampling that may disrupt analyte gas/liquid equilibrium phase distribution especially at high liquid volumes). Thus, we likely have error in the volume ratio measurements as well as measurement error of $G_C^{-1}$ of which both result in heteroskedasticity. We turn to designs for these models now.

\subsection{Optimal Heteroskedastic Design}\label{subsec:hetero_design}

Consider instead the $\errorrv_i$'s are independent, of the same functional form but now feature different variances. Note that this added assumption leads to the model explained at length by \citet{Ramachandran1996} and \citet{Asuero2007} who warn of the pitfalls of ignoring this type of structure. We now consider the generalization of Equation~\ref{eq:homo_errors},

\bneqn\label{eq:hetero_errors}
\errorrv_i \inddist e(0, \sigsq_i) \quad \text{for all $i$}.
\eneqn 

\noindent In simple linear regression with no unmeasured confounders, differing variances must be a function of $x_i$ i.e $\sigsq_i := h(x_i)$. This added complexity leaves the least squares estimator (Equation~\ref{eq:ls}) unchanged but its variance now takes into account the heterogeneity of the $\sigsq_i$'s,

\bneqn\label{eq:ls_hetero_var}
\var{\B} = \inverse{\XtX} \Xt \bOmega \X \inverse{\XtX},
\eneqn

\noindent where $\bOmega := \text{diag}(\sigsq_1, \ldots, \sigsq_n)$ i.e. the diagonal matrix with the $i$th individualied variance in the $i$th diagonal location. Equation~\ref{eq:ls_hetero_var} above can be estimated once again by a first-order Taylor approximation similar to Equation~\ref{eq:thetahat_prop_variance} (algebraic details omitted). In the heteroskedastic case, we could not find a simplification of the Taylor approximation and thus we lack an intuitive allocation method (such as Equation~\ref{eq:homo_design_alg}). For simple $h(x)$ functions, closed form solutions may exist, but we leave their exploration to future work. Instead, we resort to searching the space $\bracks{\xmin, \xmax}^n$ via numerical methods to find an allocation $x_1^\star, \ldots, x_n^\star$ which substantially lowers $\varthetahathat$ relative to naive allocations. Note that in order to perform the search, $h(x)$ must be known (or approximated) up to a constant factor.

Our software uses the Nelder-Mead numerical method \citep{Nelder1965} implemented in \texttt{R} by \citet{Bihorel2015}, a commonly used heuristic search algorithm which locates minima / maxima in many-dimensional space. As in all heuristic methods, there is no guarantee the global minimum allocation can be found, but in practice we have achieved good results especially when the algorithm searches from a number of different starting points (simulations unshown). By default, we begin with starting points of equal spacing across the interval at every allowable $\rho$ for exclusively endpoint allocation (like in the previous section) as well as a number of random starting location drawn from a multivariate uniform distribution.

If the heteroskedasticity is sufficiently strong enough, the allocation of $x_1, \ldots, x_n$ will be different from the homoskedastic design allocation of Equation~\ref{eq:homo_design_alg}. In real-world scenarios with low $n$, our experience is the heteroskedasticity is not extreme enough to substantially change the design. This is the case in our example in Section~\ref{sec:example}. Thus, this design is implemented in our software but not explored further in this paper.

\subsection{Inference for all Designs}\label{subsec:inference}

We present here a few different means for inference. We discuss approaches that construct confidence intervals of size $1 - \alpha$. Two-sided hypothesis testing of size $\alpha$ follows directly by assessing whether the null value is within the intervals; one-sided testing follows by doubling the $\alpha$ and unbounding the appropriate endpoint (left or right). Thus, we only discuss interval creation. All approaches below are implemented in the software. We defer discussion of performance and simulations to determine coverage to Section~\ref{sec:example}.

\subsubsection{The Normal Approximation}\label{subsubsec:normal_approx}

Recall the estimator $\B$ for the slope and intercept (Equation~\ref{eq:ls}). By the central limit theorem, it has an asymptotic distribution given by 

\bneqn\label{eq:asymptotically_normal}
\B \sim \multnormnot{2}{\bbeta}{\var{\B}}.
\eneqn 

\noindent Thus, the asymptotic distribution of $g(\B) := B_1 / B_0$ is given by the delta method (which is a combination of the central limit theorem in Equation~\ref{eq:asymptotically_normal} and a first order Taylor approximation as in Equation~\ref{eq:thetahat_prop_variance} line 1),

\bneqn\label{eq:delta}
g(\B) - g(\bbeta) \convd \normnot{0}{\nabla g(\bbeta)^\top ~\var{\B}~ \nabla g(\bbeta)}.
\eneqn

\noindent Thus, the interval is constructed via plugin estimates for the parameters in Equation~\ref{eq:thetahat_prop_variance}, line 2 as

\beqn
\text{CI}_{\theta, 1 - \alpha} = \bracks{\frac{b_1}{ b_0} \pm z_{\overtwo{\alpha}} \frac{s^2_e}{b_0^2} \bracks{-b_1 / b_0 ~~ 1} \XtXinv \twovec{-b_1 / b_0}{1} }
\eeqn

\noindent where $s_e^2$ is the maximum likelihood estimator for $\sigsq$. 

The analogous heteroskedastic confidence interval based on \citet{white1980}'s \qu{hetero-skedasticity-consistent covariance matrix estimator,}

\beqn
\text{CI}_{\theta, 1 - \alpha} = \bracks{\frac{b_1}{ b_0} \pm z_{\overtwo{\alpha}} \frac{1}{b_0^2} \bracks{-b_1 / b_0 ~~ 1} \inverse{\XtX} \Xt \hat{\bOmega} \X \inverse{\XtX} \twovec{-b_1 / b_0}{1} },
\eeqn

\noindent where $\hat{\bOmega}$ is the diagonal matrix of the squared residuals, is likewise inaccurate. Note that if some $x_i$'s are shared in the heteroskedastic design (which we observe in our simulations), $\hat{\bOmega}$ can feature the average squared residual pooled among common $x_i$ values. This is the method we use in our software.

Simulations (not shown) demonstrate that the asymptotic distribution does \textit{not} approximate the sampling distribution in the case of low $n$ such as the number of vials analyzed during a phase ratio measurement of $k_H$ and thus the normal approximation is not recommended for inference.

\subsubsection{Bayesian Bootstrap}\label{subsubsec:bayesian_bootstrap}

For $b = 1, \ldots, B$ times, draw $w_1, \ldots, w_n$ from a standard Dirichlet distribution (which creates a random partition of the $\zeroonecl$ interval. These weights are then used in a weighted linear regression. Record the $\thetahat^{(b)}$'s and report the endpoints of the center of the $1-\alpha$ proportion of the distribution. This produces an asymptotically valid posterior predictive interval \citep{Rubin1981} which is valid for optimal designs under both homoskedasticity and heteroskedasticity.

\subsubsection{Parametric Bootstrap for the Homoskedastic Design}\label{subsubsec:param_homo_bootstrap}

Here, we assume the error function in Equations~\ref{eq:homo_errors} and \ref{eq:hetero_errors} is now a normal density with variance unchanged. This is known to be an inaccurate assumption in HSGC as we empirically observe an error distribution with a positive skew due to the $y$ variable ($G_C^{-1}$) having a minimum of 0 which is physically imposed. Nevertheless, we are confident this method should still have reasonable coverage in real-world applications.

We employ an asymptotically valid parametric bootstrap a la \citet{Wu1986} as follows. Estimate $s^2_e$, the mean squared error of the regression (the sample variance of the residuals) and for $b = 1, \ldots, B$ times (denoted as a superscript in parentheses), simulate:

\bneqn\label{eq:param_bootstrap}
\tilde{y}_1^{(b)} = b_0 + b_1 x_1 + \tilde{e}_1^{(b)}, \ldots, ~\tilde{y}_n^{(b)} = b_0 + b_1 x_n + \tilde{e}_n^{(b)}
\eneqn

\noindent where $b_0$ and $b_1$ are the least squares estimate from the original dataset and $\tilde{e}_1^{(b)}, ~\ldots,~ \tilde{e}_n^{(b)}$ are iid sampled from $\normnot{0}{s^2_e}.$ Then, $(x_1, \tilde{y}_1^{(b)}), ~\ldots, ~(x_n, \tilde{y}_n^{(b)})$ are used to estimate $\tilde{b}_0^{(b)}$ and $\tilde{b}_1^{(b)}$. Record the $\thetahat^{(b)}$'s and report the endpoints of the center $1-\alpha$ proportion of them.

\subsubsection{A Nonparametric Bootstrap for the Homoskedastic Design}\label{subsubsec:nonparam_homo_bootstrap}

A potential weakness of the above method is its dependence on the error function being normally distributed. We can instead draw $\tilde{e}_1^{(b)}, ~\ldots,~ \tilde{e}_n^{(b)}$ as a sample with replacement from $e_1,~ \ldots,~ e_n$, the original residuals from the experimental data. Then simulate new experimental responses by setting $\tilde{y}_1^{(b)} = y_1 + \tilde{e}_1^{(b)}, ~\ldots,~ \tilde{y}_n^{(b)} = y_n + \tilde{e}_n^{(b)}$. Repeat $B$ times and record the $\thetahat^{(b)}$'s and report the endpoints of the center $1-\alpha$ proportion of them. This method is found in the original paper introducing the bootstrap \citep[Section 7]{Efron1979}.

\subsubsection{A Nonparametric Bootstrap for the Heteroskedastic Design}\label{subsubsec:nonparam_hetero_bootstrap}

Since the errors are not homoskedastic, resampling the residuals in the previous discussion will break the heteroskedastic error structure. Here, we use the method as in \citet[Section 7]{Wu1986} where the errors are multipled by $\delta \in \braces{-1,1}$ each drawn with probability 50-50. Thus, we employ Equation~\ref{eq:param_bootstrap} but here the draws of the residuals are now $\tilde{e}_1^{(b)} = \delta_1^{(b)} e_1, ~\ldots, ~\tilde{e}_n^{(b)} = \delta_n^{(b)} e_n$ where $e_1, \ldots, e_n$ are the residuals from the original data. Similar to the previous three methods, record the $\thetahat^{(b)}$'s and report the endpoints of $1-\alpha$ proportion of them. Note that \citet{Wu1986} proposes studentized residuals, a modification we do not explore here because our intution is that its performance improvement would be small.

\subsubsection{A Note about the Standard Non-Parametric Bootstrap}

Note that the sample sizes common in a laboratory setting (on the order of 10) are relatively small. The standard row-resampling non-parametric bootstrap (for $B$ runs draw $m \leq n$ rows with replacement and run analysis) would be problematic for two reasons. (1) Its convergence is slow and since our sample sizes are small we are thus not confident in its convergence. (2) We frequently have one or two $x$ values at one of the endpoints and a bootstrap sample with replacement may not choose those points. This would yield estimates for the slope and intercept that do not exist. As such, many of the bootstrap samples would must be discarded. Further, the theory for properly handling such illegal bootstrap samples is not clear. Thus, we do not explore this method here nor implement it in our software package.

\section{An Example: The $k_H$ of Napthalene in Water}\label{sec:example}

To evaluate the design approach developed herein, we measured the $k_H$ value for Napthalene. This common VOC is a widely recognized persistant pollutant in the environment whose Henry's coefficient of 0.053 according to the \citet{EPACalc} falls within the middle of the range of those commonly being evaluated with PRV. We now describe our experimental setup, then walk through the experimental design and estimation and finally provide confidence intervals for our estimate.

\subsection{Preparation}

Napthalene was mixed in a pure aqueous solution was measured at 40$^\circ$C. A stock solution (3.8L) of napthalene at approximately 1 to 10\% of its aqueous solubility at 20$^\circ$C have been prepared using purified water (18.2m$\Omega$, UV disinfected, 0.2$\mu$m filtered, Flex3 by Elga). For each solution, twenty vials were prepared at each of the following liquid volumes: 1.1, 1.3, 1.5, 1.8, 2.3, 3.2, 5, 7.5, and 15ml. The quantities of these solutions were gravimetrically determined using an analytical balance (Mettler Toledo, XS205 $\pm$ 0.01mg) and the volumes in solution were adjusted for temperature at the point of analysis. This spacing of liquid volumes were chosen to provide a more equal spacing of ratio values (the $x$ variable) and the wide range was chosen for accuracy.

The sample vials and their gas phase were temperature and phase equilibrated, sampled and analyzed using a HSGC. The instrument used was from Thermo Scientific Tracer 1310 GC with a headspace autosampler (RSH) equipped with a heated agitator and the detector was a single quadrupole mass spectrometer detector (ISQ). For this study, the vials were agitated and incubated at the target temperature for 1 hr. Subsequently, 0.3 ml of headspace was extracted with 0.2 ml being injected into a gas chromatograph with a split ratio of 20 to 1 for analysis. High purity helium was used as the carrier gas at a flowrate of 1.00ml/min. We now design and run the experiment.

\subsection{Computing our Design}\label{subsec:homo_design_example}

First, we gathered calibration data which was the 80 vials ran at the solution volumes listed in the previous section. We used this data to determine $\xmin$ and $\xmax$ and an estimate of $\sigma$, the standard error of measurement (see Equation~\ref{eq:homo_errors}).

We first run the \textit{trial run} described in the previous section. From this data, we notice that liquid volumes below 1.5mL and above 5mL yield non-linear relationships in $G_C^{-1}$ versus the gas/liquid ratio. In the case of individual vials, those with the lowest liquid volumes would likely experience the greatest relative loss of solute and solvent (if it is highly volatile) during sample preparation.  The quality of the data from the larger volume liquid vials may have analagous issues.  The gas concentration is the highest of the set in those vials and thus most tests the assumption of linearity in the GC response. 

Thus, our experimental domain is between $\xmin = 0.33$ and $\xmax = 14.44$. Note that in a real application, this initial run would not be performed. It is assumed the experimenter has the $x$ range fixed from previous experience. Here we also assume $n=10$ because this is a reasonable number of vials to run. Recall from Equation~\ref{eq:thetahat_prop_variance} that we must begin with a guess of the $k_H$ value, $\theta_0$. Here, we employ a guess of the true value 0.053. We analyze robustness to this guess

Given the numbers above, $\xmin,~\xmax,~n$ and $\theta_0$, we compute the optimal design given by Equation~\ref{eq:homo_optimal_design}. In our \texttt{R} software package, this is performed via

\begin{verbatim}
> oed_for_slope_over_intercept(n = 10, 
    xmin = 5/15, xmax = 19/1, theta0 = 0.053)
\end{verbatim}

\noindent which results in 7 $\xmin$'s and 3 $\xmax$'s for the design. 

\subsection{Our Design Performance}\label{subsec:design_performance}

\begin{figure}[h]
\centering
\centering
\includegraphics[width=4.3in]{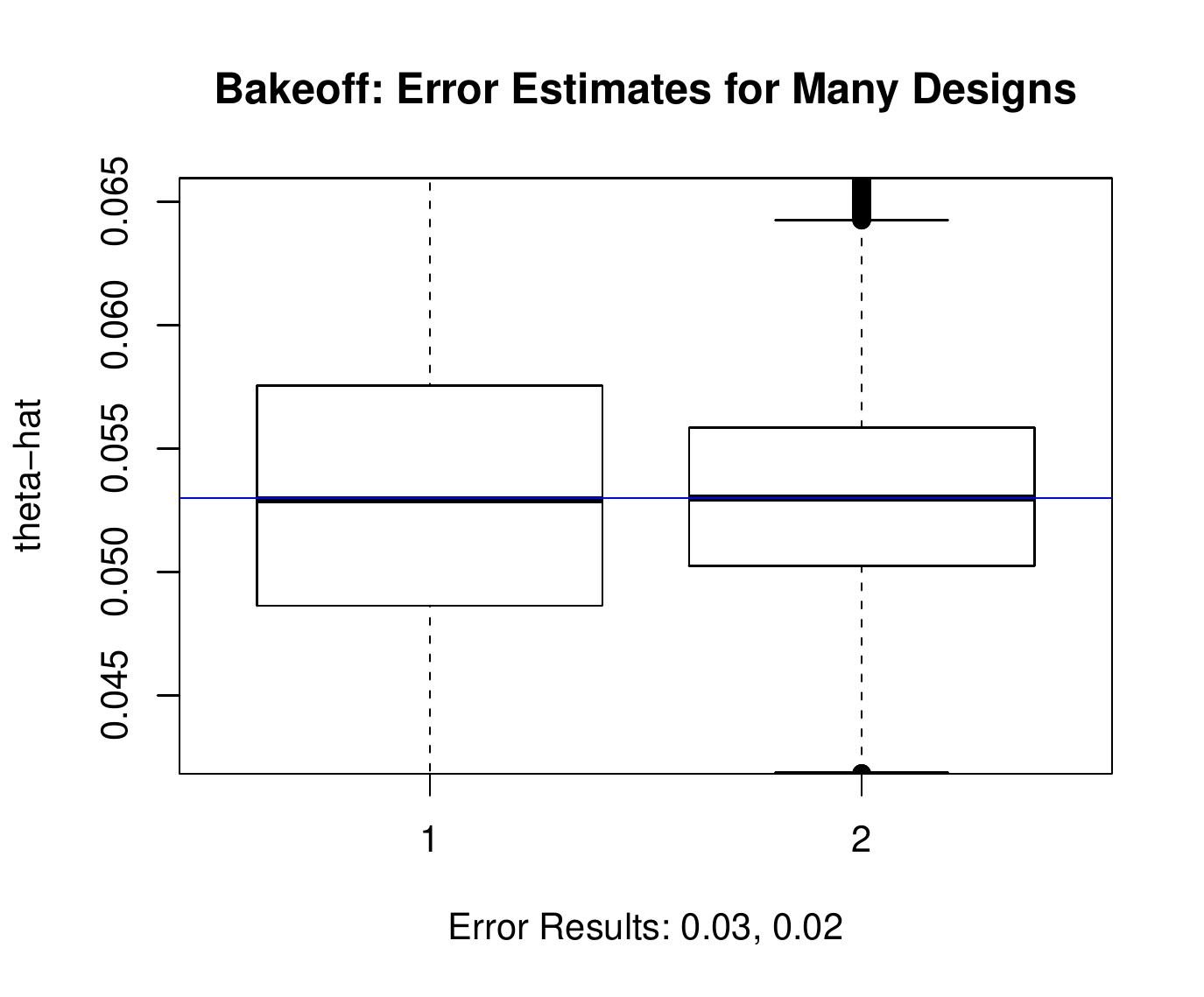}
\caption{A box-and-whisker plot for 50,000 simulations comparing the $\thetahat$ empirical distributions of the two designs simulated under the same homoskedastic linear response model (see text for details). The horizontal line indicates the true value of $\theta = 0.5$. Error estimates (as computed by the intercentile range --- the 99\%ile - 1\%ile) of the empircal distributions are printed below the $x$-axis label. This plot is generated via the function \texttt{design\_bakeoff} in our software.}
\label{fig:naive_vs_optimal}
\end{figure}

Our claim is this allocation constitutes an \qu{optimal design.} But what is the gain over alternative designs with different gas:liquid selections? Figure~\ref{fig:naive_vs_optimal} displays the result of a simulation of two designs: (1) even spacing allocation from $\xmin$ to $\xmax$ and (2) our proposed optimal design. We simulate under the response model where $\theta_0 = \theta = 0.5$ (i.e. as if the experimenter's a priori guess was the correct value). We will see that our proposed design here is very robust if the $\theta_0$ guess was not correct in Section~\ref{subsec:robustness}. 

In order to perform this simulation, we need guesses of two nuisance parameters, the true intercept $\beta_0$ and the true standard error of the residuals, $\sigma$ (see Equation~\ref{eq:homo_errors}). We record an estimate of $\sigma \approx 3.2 \times 10^{-10}$ from the trial run and $\beta_0 = 3.9 \times 10^{-9}$.  Note that once again, this initial run would not be performed. Thus, in order to gauge efficiency of the allocation we propose vis-a-vis others, one can guess $\beta_0$ and $\sigma$ to be sets of reasonable previous values.

Note that all designs are unbiased when examining the median of the $\thetahat$ distributions. But the spreads are wildly different. When measured by the intercentile range (the 99\%ile minus the 1\%ile i.e. the center 98\% of the distribution), the  naive estimate yields an error 61\% more than our optimal design estimate. The gain is the same when measured by standard error and the relative efficiency (as measured by the ratio of the variance of the naive to the optimal) is 2.6. Confidence intervals for the true $k_H$ are 60\% wider under the naive allocation versus the proposed optimal allocation. Note that these figures are unique to napthalene and our HSGC machinery. Much higher gains may be seen in other VOCs and other equipment where the linear relationship is measured with higher noise ($\sigma$).


This enhanced performance represents a marked improvement for PRV measurements and thus the confidence with which its findings can be relied upon.

\subsection{Examining Robustness of our $k_H$ Guess}\label{subsec:robustness}

We now return to what was seemingly a weakness in our design. We required a guess, $\theta_0$. How robust is this optimal design to deviations from this initial guess? And how robust are the impressive gains in efficiency? These two questions can be investigated by imagining if the real $\theta$ was different and simulating the efficiency loss across a large range of realistic $\theta$'s. 

\begin{figure}[h]
\centering
\includegraphics[width=6in]{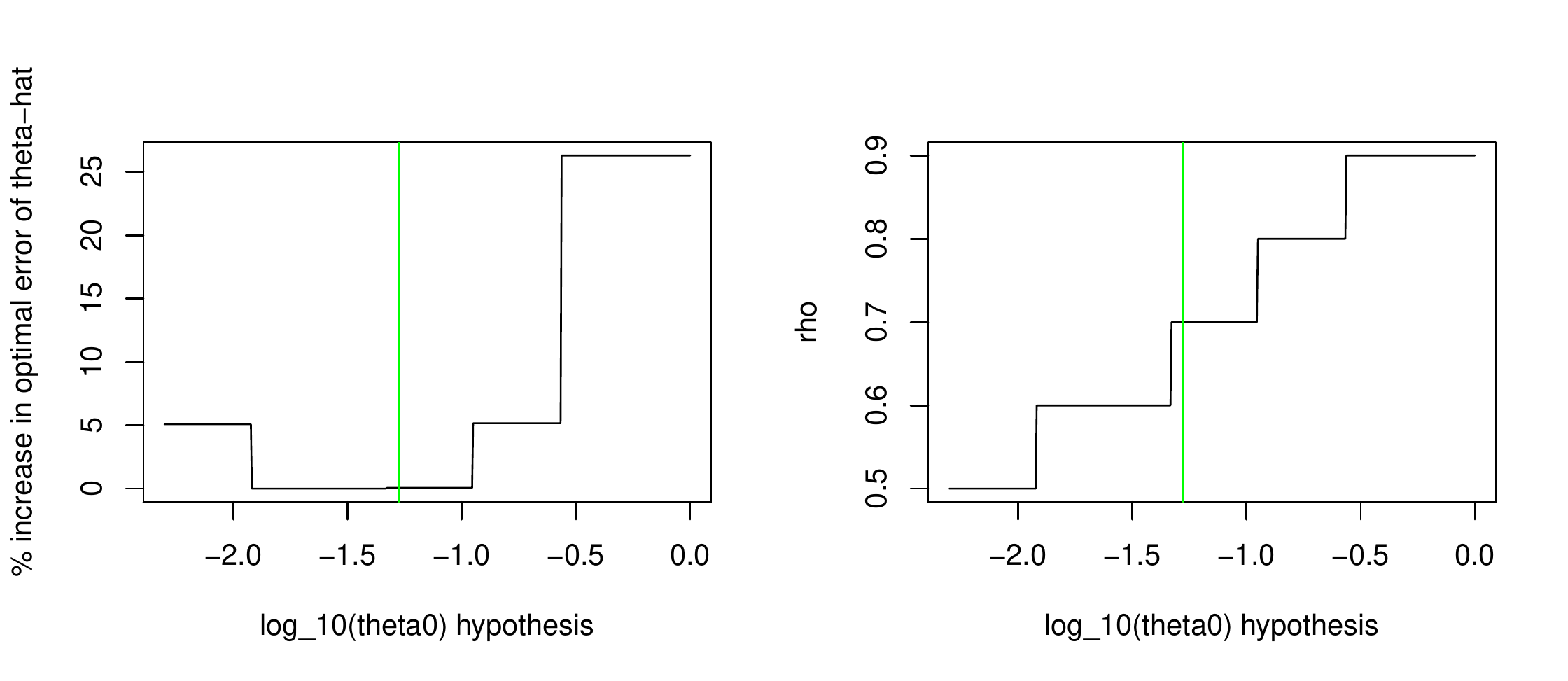}
\caption{The left plot illustrates estimated error percentage increase (as measured by intercentile range) versus the minimum when $\thetahat$ is estimated with incorrectly specified designs (computed via the method described in Section~\ref{subsec:homo_design}). The $x$-axis varies over a wide range of possible $\theta_0$'s. The right plots show the value of $\rho^\star$ over the same $\theta_0$ range. The green line is the error estimate for the design optimized for the true $\theta = 0.053$ and thus respresents the minimum error. Note that $\xmin,~\xmax$ and $n$ are set to the values in the body of the paper and the simulation uses the values from the trial run. This plot is generated via the function \texttt{err\_vs\_theta0\_plot\_for\_homo\_design} in our software. The plot's values represent the average of 50,000 simulations for each of the 500 positions on the $x$-axis.}
\label{fig:robust_to_other_thetas_napth}
\end{figure}

We simulated estimator error (intercentile range) for $\theta = 0.053$ . Examining Figure~\ref{fig:robust_to_other_thetas_napth}, if $\theta$ was truly 0.053, it seems any value of $\theta_0$ between 0.012 and 0.11 would yield about the same performance. If $\theta_0$ was lower, the most performance will suffer is about 5\%. 
And if $\theta_0$ was higher, even up to 1 (which is 20 times the true $k_H$ value), we expect only about 25\% more error. We would like to stress that these error differences (even in the worst case) are substantially lower than the $\approx$60\% if the naive design was employed (see third paragraph in the previous section). Thus, it appears our design is robust to incorrect guesses of the true parameter --- the practitioner need only be in the right ballpark for our design to be valuable.

Some potential applications could be difficult to project a $\theta_0$. For instance, consider water from a fracking well and the level of suspended drilling mud and salts in solution to be unknown and variable temperatur, then one may have a difficulty estimating $k_H$.  Such a scenario is common in hydraulic fracturing for natural gas and oil recovery. Thus, this robustness check can be important and is readily available in the software. 

\subsection{$k_H$ Inference Performance}\label{subsec:inference_example}

How do these confidence intervals perform in practice in the homoskedastic case? We simulate under the design in Section~\ref{subsec:homo_design_example} and assume homoskedasticity. Table~\ref{tab:homo_cis} shows 95\% confidence intervals for the optimal and the naive design for each interval construction method detailed in Section~\ref{subsec:inference} appropriate for homoskedastic data. 

\begin{table}[htp]
\centering
\begin{tabular}{lcc}

Interval Method Name  & Example   & Estimated \\
(section in text) &  Interval &  Coverage \\ \hline

Normal Approximation && \\
Homoskedastic (\ref{subsubsec:normal_approx})  &   [0.0477, 0.0478]  & 0.007 \\
Normal Approximation && \\
Heteroskedastic (\ref{subsubsec:normal_approx})  & [0.0477, 0.0478]  & 0.006 \\
Bayesian && \\
Weighting (\ref{subsubsec:bayesian_bootstrap}) & [0.0427, 0.0530] & 0.851 \\ 
Parametric &&\\
Homoskedastic (\ref{subsubsec:param_homo_bootstrap}) &  [0.0418, 0.0545]  & 0.920 \\ 
Nonparametric &&\\
Homoskedastic (\ref{subsubsec:nonparam_homo_bootstrap}) &  [0.0424, 0.0541] & 0.884 \\
Nonparametric &&\\
Heteroskedastic (\ref{subsubsec:nonparam_hetero_bootstrap}) &  [0.0422, 0.0539]  & 0.887 \\
\hline
\end{tabular}
\caption{Example 95\% confidence intervals for one simulated homoskedastic dataset (described in the text) with the true $\theta = 0.053$ for each interval construction method as well as approximate coverage determined by tabulating 1,000 simulations.}
\label{tab:homo_cis}
\end{table}

Note that intervals constructed based on the asymptotic normal approximation (rows 1 and 2) have poor coverage. Once again, $n=10$ is small and the tails on the finite-small $\thetahat$ density are very much thicker than Gaussian tails. The resampling techniques perform much better and come close to the 95\% coverage. Once again, these are asymptotically valid interval construction methods and there is no guarantee of exact performance in small samples. Interesting to note is the last method which is built for heteroskedasticity does not perform terribly. This is in line with the general image that emerged during design: barring a huge sample size or large heteroskedasticity, the homoskedastic design and interval methods perform robustly. In unshown simulations, we have observed that the heteroskedastic intervals have fickle performance but the homoskedastic intervals are more robust.

\section{Efficiency Gains in Past Measurements}\label{sec:past_examples}

We have demonstrated gains in efficiency of about 60\% versus equal spacing when using our proposed optimal design when measuring the $k_H$ of napthalene in water. As we mentioned, measuring the $k_H$ via PRV method is quite widespread. Our goal here is to demonstrate efficiency gains on some previous measurement efforts via reasonable simulation.

Table~\ref{tab:previous_studies_setups} gives the experimental specifications of the previous studies (for the citation of the numbered study, cross-reference in Table~\ref{tab:previous_studies}) along with results (the loss column).

\begin{table}[htp]
\centering
\begin{tabular}{l|cccccc}
   & Vial Volume & Liquid Volumes & Gas-liquid  & \#  & &  loss  \\
\#   & (mL) & (mL) & ratios & reps & $n$ &  (\%) \\\hline
1& 20.0 & 0.06, 0.08, 0.12, 0.24     & 80.9, 161.8, 242.8, 323.7                                       & 3 & 12 & 49\\
2& 20.0 & 0.02, 0.04, 0.1, 0.5              & 39 ,199, 499, 999                                     & 3 & 12 & 42 \\
3& 20.2 & 2, 5, 10                            & 1.01, 3.03, 9.08                                    & 3 & 9 & 18\\
4& 22.4 & 1, 2, 3, 4                           & 4.6, 6.5, 10.2, 21.4                                    & 2 & 8 & 24\\
5& 22.3 & 0.1, 0.3, 0.5, 1, 2               & 10.2, 21.3, 43.6, 73.3, 222                                 & 2 & 8 & 23\\
6& 22.4 & 0.05, 0.2, 0.5, 2               & 10.2, 43.8, 111, 447                                & 3 & 12 & 26\\
7& 22.0 & 1, 2, 3, 4              & 4.5, 6.3, 10, 21                             & 3 & 12 & 22\\
8& 20.0 & 0.05, 0.1, 1, 10              & 39, 199, 499, 999                                     & 1 & 4 & 29\\
9& 22.0 & 1.5, 2, 2.5, & 13.7, 10.0, 7.8, 6.3, & 3 & 24 & 37 \\
 &         & 3, 4, 5, 7, 10 & 4.5, 3.4, 2.1, 1.2 & & & \\
10& 22.3 & 1, 2, 3, 4               & 4.6, 6.4, 10.2, 21.3                                 & 1 & 4 & 23 \\
\hline
\end{tabular} 
\caption{Experimental PRV setups of ten previous studies. Replicates indicated repeated measurements on the same gas-liquid ratio. Last column indicates the simulated increase in standard error of the $k_H$ estimate which is expected when using the designs in the previous studies over the optimal design proposed here (10,000 simulations).}
\label{tab:previous_studies_setups}
\end{table}

To simulate efficiency gain, we use the $\beta_0$ and $\sigma$ estimates from our napthalene example found in Section~\ref{subsec:design_performance}. We then compare the designs used in the original study i.e. the gas-liquid ratios found in column 3 of Table~\ref{tab:previous_studies_setups} to the optimal designs found by Equation~\ref
{eq:homo_optimal_design} using the $n$'s from column 5. Note that we assume also the operating range defined by the minimum and maximum values of the ratios in column 3. Since our design is mostly robust to changes in the guess of the true $k_H$ (see Section~\ref{subsec:robustness}), we set $\theta_0 = 1$ for all runs regardless of the VOC. A more careful simulation would involve estimating all five measurement-specific parameters ($\xmin$, $\xmax$, $\beta_0$, $\sigma$ and $\theta_0$). Thus, our results are conservative; with better prior ideas of $k_H$, our method will have stronger improvements.

The last column (loss \%) displays the increase in standard error of the $k_H$ estimate that can be expected given the designs used in the study over the optimal designs which is on average about 30\%. And with noisier HSGC measurements, the gain can be substantially more since the estimator has infinite moments. 100\%, 200\% and even 1000\% increases in standard error are not uncommon in simulations.

\section{Discussion}\label{sec:discussion}

We have presented a means to derive optimal experimental designs when the parameter of interest is the slope to intercept ratio such as the case in HSGC to measure Henry's Constant. Using our designs, these measurements become far more accurate in comparison to naive designs. How much more accurate? Since the tails of the estimator created by dividing the sample slope by the sample intercept are large, it is not possible to say how much more accurate when considering the canonical metric of accuracy, standard error.

We have also shown in this work that our design is fairly robust to misspecification of both the guess of the parameter which is required a priori as well as failing to specify the heteroskedasticity in the response. Misspecification in the parameter estimate is assessed easily with routines in our provided open-source software suite.

The application showcased in this paper was the measurement of the $k_H$ of napthalene using the phase ratio method whose parameter of interest was the slope-to-intercept ratio. Here, naive allocations performed about 60\% worse than our optimal design when measured by standard error. We then simulated based on designs in PRV measurements of previous studies going back to 1993 and found similar order improvements.

Applications which share interest in the slope divided by the intercept abound in the literature expecially in experimental physical chemistry and is well known in textbooks in the field. We list a few examples of applications here: measuring the quenching of singlet oxygen \citep{Foote1970}, investigating electronic field dependence \citep{Chance1973}, measuring the dissociation constants of enzyme-substrate \citep{Strickland1975, Pollegioni1994} and measuring rate constants to determine acidity \citep{Czapski1999}. There are also a number of applications in social science such as a metric which aid in estimating cause-specific mortality of adult female elk \citep{Brodie2013}. The designs described in this paper are immediately applicable to practitioners in these fields.

\subsection{Future Work}\label{subsec:future_work}

For a better approximation of the variance of $\thetahat$ (Equation~\ref{eq:thetahat_prop_variance} line 1), we can use a higher order Taylor series approximation. This would also improve the confidence intervals discussed in Section~\ref{subsubsec:normal_approx}. It is our intuition that for our low $n$ situation, that such an approximation will not make a large difference. 

Further, the estimators for both the slope and intercept are asymptotically normal (Equation~\ref{eq:asymptotically_normal}) with a fast rate of convergence since the $e$ function in our measurement device (see Equations~\ref{eq:homo_errors} and \ref{eq:hetero_errors}) is well-behaved. Thus, $\thetahat$ is a ratio of two correlated normal random variables. The exact density thereof has been found in closed form dating back to \citet{Fieller1932}. \citet[Equation 3]{Hinkley1969} gives its cumulative distribution function $F_{\thetahat}$ as a function of the standard bivariate normal integral. \citet[Equation 13]{Pham-gia2006} gives its density as a function of Kummer's classic confluent hypergeometric function. However, all expressions in the literature demand knowledge of the true means and variances of numerator and denominator as well as the correlation (i.e., in our setup, $\beta_0,~\beta_1$ and $\sigsq$). Nevertheless, it may be possible that employing guesses for the parameters then minimizing the quantity $F_{\thetahat}((1-\alpha) / 2) - F_{\thetahat}(\alpha / 2)$ for some $\alpha$ as a function of the $x_i$'s may be a fruitful path towards further optimizing the design we present here. Heteroskedastic models would be more cumbersome. Using plug-in estimators may also be a strategy for building asymptotically valid confidence intervals (Section~\ref{subsec:inference}). To our knowledge, a density function for the random variable parameterized by estimates of the means and covariances from the data (i.e. $b_0,~b_1$ and $s^2_e$ in our application) has not yet been discovered. Plugging in those estimates in lieu of the true parameter values (especially at low $n$) would yield inaccurate figures, so we have left this method out of this manuscript.

When measuring $k_H$, we know that $\theta$ is always positive and $\beta_0$ is always positive. Thus, negative estimates and estimates close to zero should be shrunk to a more reasonable positive value. Thus, a Bayesian model can be employed such as

\beqn
\underbrace{\mathbb{P}(\theta~|~\thetahat,~\X)}_{\text{posterior}} ~\propto~ \underbrace{\mathbb{P}(\thetahat~|~\theta,~\beta_0,~\sigsq,~\X)}_{\text{likelihood}}~ \underbrace{\cprob{\theta}{\beta_0,~\sigsq,~\X}}_{\theta ~\text{prior}} ~\underbrace{\cprob{\beta_0}{\sigsq,~\X}}_{\beta_0 ~\text{prior}}~ \underbrace{\cprob{\sigsq}{\X}}_{\sigsq~ \text{prior}}.
\eeqn

\noindent The likelihood can be specified using the exact density of the ratio of two correlated normal random variables (which is now possible in the Bayesian framework since we assume the parameters are known) or the asymptotic normal which leads to possible conjugacy. The prior on $\theta$ can be constructed with previous measurements from the application of interest. For example, the bulk of $k_H$ values are between 0.1 and 1.5. Likewise the prior on $\beta_0$ can be estimated from van't Hoff plots \citep{Mackay2006} and the $\sigsq$ prior can be an inverse-gamma (the typical prior on $\sigsq$ in a Bayesian model) with hyperparameters set based on previous experience with the measurement device. A heteroskedastic extension follows with a bit more specification.

Here, the optimal design can be found by minimizing the posterior variance as a function of the $x_i$'s. Contrary to the frequentist setup, now an estimate of $\thetahat$ is needed (the prior expectation is a reasonable place to start). Without conjugacy between the likelihood and priors, this would be computational task but not overly burdensome given modern sampling techniques. Our intuition is the design would not be too different from the designs found in Sections~\ref{subsec:homo_design} and \ref{subsec:hetero_design}.

But this method is worthwhile in that after the experiment, the posterior distribution can be estimated via numerical sampling, estimates of $\theta$ can be found by computing the posterior mean (or other methods) and intervals can be constructed by computing quantiles of the posterior distribution (see \citealp{Gelman2014} for an introduction to these methods). We would expect the Bayesian estimate and intervals to perform much better than the frequentist methods suggested in this paper especially at small values of $\theta$ and when $\beta_0 \approx 0$. However, for values significantly larger, our intuition is the improvement would be small especially at sample sizes employed in lab experimentation. For instance, at $\theta = 0.053$, Figure~\ref{fig:naive_vs_optimal} demonstrates that not one in 50,000 estimates of $\thetahat$ are illegally below 0.

Further, the interval construction methods of Section~\ref{subsec:inference} can be further explored to determine coverage in different situations. Those that involve bootstrap resampling can be improved by employing the work of \citet{Efron1987} for instance.

And lastly, it would not be of great difficulty to study, as a new parameter of interest, the intercept divided by the slope. The software can be trivially changed as well to make this accomodation.

\subsection*{Replication}

The figures, tables and computations found in this paper can be replicated by running \texttt{paper\_duplication.R} found in the github repository for the \texttt{optDesignSlopeInt} package at \url{https://github.com/kapelner/optDesignSlopeInt} which is open source under GPL2.

\bibliographystyle{apalike}
\bibliography{oed_paper}

\appendix

\section{Proof of the Absence of Interior Points}\label{app:proofs}

Assume the existence of one interior point $y$ in the optimal design for the homoskedastic response model. Consider $a$ points at $\xmin$ and thus $n-a-1$ points at $\xmax$. Let $z := \xmax - y$ and $\delta := \xmax - \xmin$. Thus,

\beqn
\xbar(z) &=& \overn{a \xmin + (n-a)\xmax - z} \quad \text{and}\\
s^2_x(z) &=& (n-1) \parens{a \squared{\xmin - \xbar} + (n-a) \squared{\xmax - \xbar} + \squared{\xmax - z - \xbar} - \squared{\xmax - \xbar}}
\eeqn

\noindent where some algebra gives the following simplification

\beqn
s^2_x(z) = \overn{z^2 (n-1) - 2az\delta + a(n-a)\delta^2}.
\eeqn

\noindent Recall that we wish to minimize the expression in Equation~\ref{eq:thetahat_prop_variance}, line 3,

\beqn
\varthetahathat \propto \frac{\squared{1 + \theta_0 \xbar(z)}}{s^2_x(z)} \propto \frac{\squared{1 + \theta_0 \xbar(z)}}{s_{xx}(z)} .
\eeqn

\noindent where $s_{xx}(z) := (n-1) s^2_x(z)$. Using some of the substitutions above and taking the derivative we find

\beqn
h(z) = \frac{-2 s_{xx} (1 + \theta_0 \xbar) \theta_0 - 2 \squared{1+\theta_0 \xbar} \parens{z \parens{n-1} - a\delta})}{n s_{xx}}
\eeqn

\noindent and since we are only concerned with the sign of the derivative it suffices to consider

\beqn
h^*(z) := -\parens{s_{xx}(z) \theta_0  + (1 + \theta_0 \xbar(z))(z(n-1) - a\delta)}.
\eeqn

\noindent Substituting once again for $\xbar$ and $s_{xx}$, we find that $h^*(z)$ is a linear function of $z$ with the linear coefficient being

\beqn
n - 1 + \theta_0 \parens{(n-1) \xmax - a \delta}
\eeqn

\noindent which is always greater than zero since $a \leq n - 1$ and $\delta \leq \xmax$. Hence, $h'(z)$ can be zero at most once; we denote this point $z_0$. The above analysis shows that $h'(z) > 0$ for $z < z_0$ and $h'(z) < 0$ for $z > z_0$ so $z_0$ is a local maximum. 

Thus, to minimize $\varthetahathat$, the optimal solution is to set $y$, the one possible interior point, to be equal to $\xmin$ or $\xmax$. $\blacksquare$

\end{document}